\documentclass[aps,prb,floatfix,twocolumn,showpacs,reprint,superscriptaddress]{revtex4-1}
\usepackage{amsfonts}
\usepackage{mathrsfs}
\usepackage{amsmath}
\usepackage{color}
\usepackage{graphicx}
\usepackage{bm}
\usepackage{amssymb}
\usepackage{xspace}
\usepackage{epstopdf}
\usepackage{dcolumn}
\usepackage{longtable}
\usepackage{multirow}
\usepackage[colorlinks=true, letterpaper=true, pdfstartview=FitV, linkcolor=blue, citecolor=blue, urlcolor=blue]{hyperref}

\begin{document}

\title{Strain effects on electronic and optic properties of monolayer C$_2$N holey two-dimensional crystals}

\author{Shan Guan}
\affiliation{Research Laboratory for Quantum Materials, Singapore University of Technology and Design, Singapore 487372, Singapore}
\affiliation{School of Physics, Beijing Institute of Technology, Beijing 100081, China}

\author{Yingchun Cheng}
\affiliation{Institute of Advanced Materials (IAM), Nanjing Tech University, Nanjing 211816, China}

\author{Chang Liu}
\affiliation{School of Physics, Beijing Institute of Technology, Beijing 100081, China}

\author{Junfeng Han}
\affiliation{School of Physics, Beijing Institute of Technology, Beijing 100081, China}

\author{Yunhao Lu}
\affiliation{School of Materials Science and Engineering, Zhejiang University, Hangzhou 310027, China}

\author{Shengyuan A. Yang}
\email{shengyuan\_yang@sutd.edu.sg}
\affiliation{Research Laboratory for Quantum Materials, Singapore University of Technology and Design, Singapore 487372, Singapore}

\author{Yugui Yao}
\email{ygyao@bit.edu.cn}
\affiliation{School of Physics, Beijing Institute of Technology, Beijing 100081, China}

\begin{abstract}
A new two-dimensional material, the C$_2$N holey 2D (C$_2$N-$h$2D) crystal, has recently been synthesized. Here we investigate the strain effects on the properties of this new material by first-principles calculations. We show that the material is quite soft with a small stiffness constant and can sustain large strains $\geq 12\%$. It remains a direct gap semiconductor under strain and the bandgap size can be tuned in a wide range as large as 1 eV. Interestingly, for biaxial strain, a band crossing effect occurs at the valence band maximum close to a 8\% strain, leading to a dramatic increase of the hole effective mass. Strong optical absorption can be achieved by strain tuning with absorption coefficient $\sim10^6$ cm$^{-1}$ covering a wide spectrum. Our findings suggest the great potential of strain-engineered C$_2$N-$h$2D in electronic and optoelectronic device applications.
\end{abstract}

\maketitle

Since the discovery of graphene, two-dimensional (2D) materials have attracted tremendous interest due to their many fascinating properties.\cite{novo2004,geim2007,xu2013,butl2013} One current focus is to explore new 2D materials with suitable semiconducting bandgaps for device applications. Recently, such a new 2D crystal, C$_2$N holey 2D (C$_2$N-$h$2D) crystal, has been successfully synthesized.\cite{mahm2015} The material has a direct bandgap (with reported optical gap size around $2$ eV), and a field-effect-transistor fabricated based on it shows a high on/off ratio of $10^7$, suggesting its great potential for applications in electronics and optoelectronics.\cite{mahm2015,zhang2015,zhang2015b,xu2015}

For application purposes, it is crucial to have the ability to tailor electronic properties of the material. Strain has long been known as an effective mechanism for tuning properties of semiconductors. It is especially useful for low-dimensional systems since they can usually sustain much larger strains than their bulk crystals. In particular, it has been shown that 2D materials, such as graphene, MoS$_2$, and phosphorene, have excellent mechanical flexibility (with critical strains $\geq 25\%$),\cite{kim2009,lee2008,cast2012,bert2011,peng2014} which makes strain an extremely powerful approach for engineering the properties of 2D materials.\cite{levy2010,guin2010,feng2012,zhang2013a,rodi2014,fei2014}

Motivated by the urgent need in understanding the physical properties of the newly discovered C$_2$N-$h$2D material and by the great interest in engineering it for applications, in this work, we investigate the effects of biaxial and uniaxial strains on the electronic and optic properties of monolayer C$_2$N-$h$2D crystals using first-principles calculations. We find that the material is quite flexible with a small stiffness constant and can withstand strains $\geq 12\%$. Under different types of strain, while still maintaining a direct bandgap, the gap size can be tuned in a wide range as large as 1 eV. More interestingly, for biaxial strain, due to different bonding characters of the bands, there is a switch of band ordering near the valence band maximum (VBM) at a critical strain $<8\%$, leading to a strain-induced dramatic increase of the hole effective mass. Despite its atomic thickness, this material shows fairly large optical absorption over most visible light spectrum, and the absorption profile as well as the peak positions can be effectively tuned by strain.

\begin{figure}
\includegraphics[width=9cm]{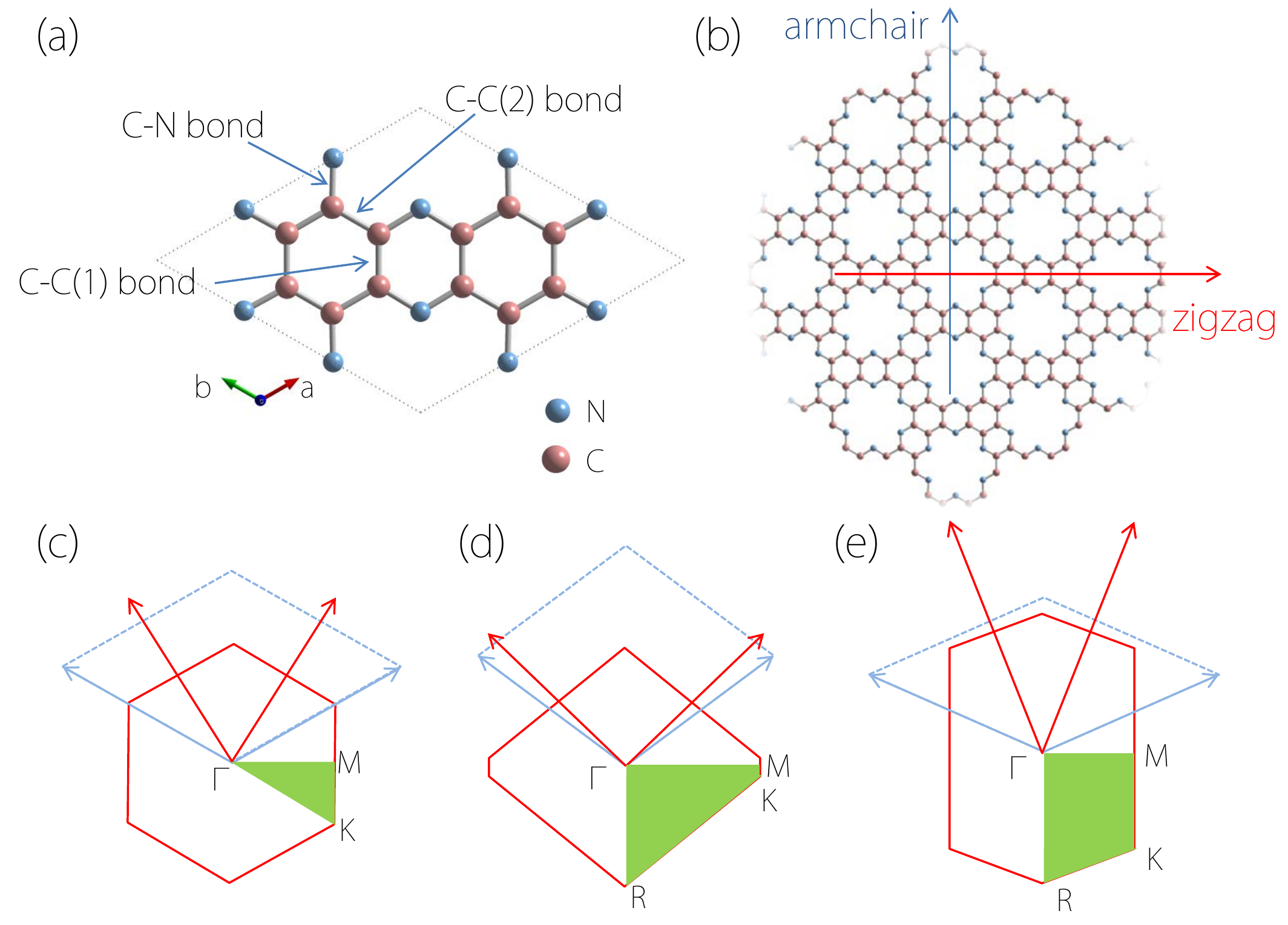}
\caption{(a) Unit cell of monolayer C$_2$N-$h$2D crystal structure. (b) Illustration of armchair and zigzag directions. (c-e) schematically show the unit cells (in blue), the Brillouin zones (in red), and the irreducible
Brillouin zones (filled with green color) corresponding to (c) biaxial strain, and uniaxial tensile strain along (d) armchair or (e) zigzag direction. High symmetry points are labeled.}
\label{fig1}
\end{figure}

The first-principles calculations were performed based on the density functional theory (DFT) as implemented in the Vienna \emph{ab-initio} simulation package.\cite{vasp1,vasp2} The projector augmented wave (PAW) potentials were used to treat the core electrons.\cite{paw1,paw2} The crystal lattice geometry was optimized using the Perdew-Burke-Ernzerhof (PBE) functional.\cite{pbe} For semiconductors, in order to obtain a more accurate description of the electronic states, the band structures were calculated using the hybrid functional (HSE06).\cite{hse1,hse2} The plane-wave energy cutoff was set to be 520 eV, and $\Gamma$-centered $k$-point meshes of sizes $9\times 9\times 1$ and $11\times 11\times 1$ were used for the geometry optimization and static electronic structure calculations, respectively. The thickness of the vacuum region was taken to be at least 16 {\AA} to avoid artificial interactions between the layer and its periodic images, and the results for stress and dielectric functions are suitably renormalized to exclude the vacuum region by using the experimental interlayer distance of 3.28 \AA\; as the effective thickness.\cite{mahm2015}  The convergence criteria for energy and force were set to be $10^{-6}$ eV and 0.005 eV/{\AA}, respectively.

As shown in Figs.~\ref{fig1}(a) and \ref{fig1}(b), the monolayer C$_2$N-$h$2D crystal is a single 2D sheet of atoms with uniform periodic holes in a fused aromatic network structure. It can be viewed as a 2D honeycomb lattice of benzene rings connected through nitrogen atoms. From geometry optimization, we find that the lattice constant is $8.330$ \AA, and the bond lengths are 1.337 \AA, 1.429 \AA, and 1.470 \AA\; for the C-N bond, the C-C(1) bond, and the C-C(2) bond, respectively (see Fig.~\ref{fig1}(a)). The C-C(2) bonds, which face the holes, are about 3\% longer than the C-C(1) bonds. Hence each benzene ring is slightly distorted due to the surrounding N atoms. Our results agree well with the experimental values with a lattice constant $\simeq8.30$ \AA.\cite{mahm2015} Starting from the fully relaxed structure, we consider in this work three types of strain, namely, the biaxial strain and two uniaxial strains along the zigzag and armchair directions, as indicated in Fig.~\ref{fig1}(b). Since 2D materials are prone to wrinkle under lateral compression, we will focus on tensile strains here, which are defined as $\varepsilon=(\ell-\ell_0)/\ell_0\times100\%$ where $\ell$ and $\ell_0$ are the lengths of the cell (along the strain direction) of the strained and original structures, respectively.

\begin{figure}
\includegraphics[width=8.6cm]{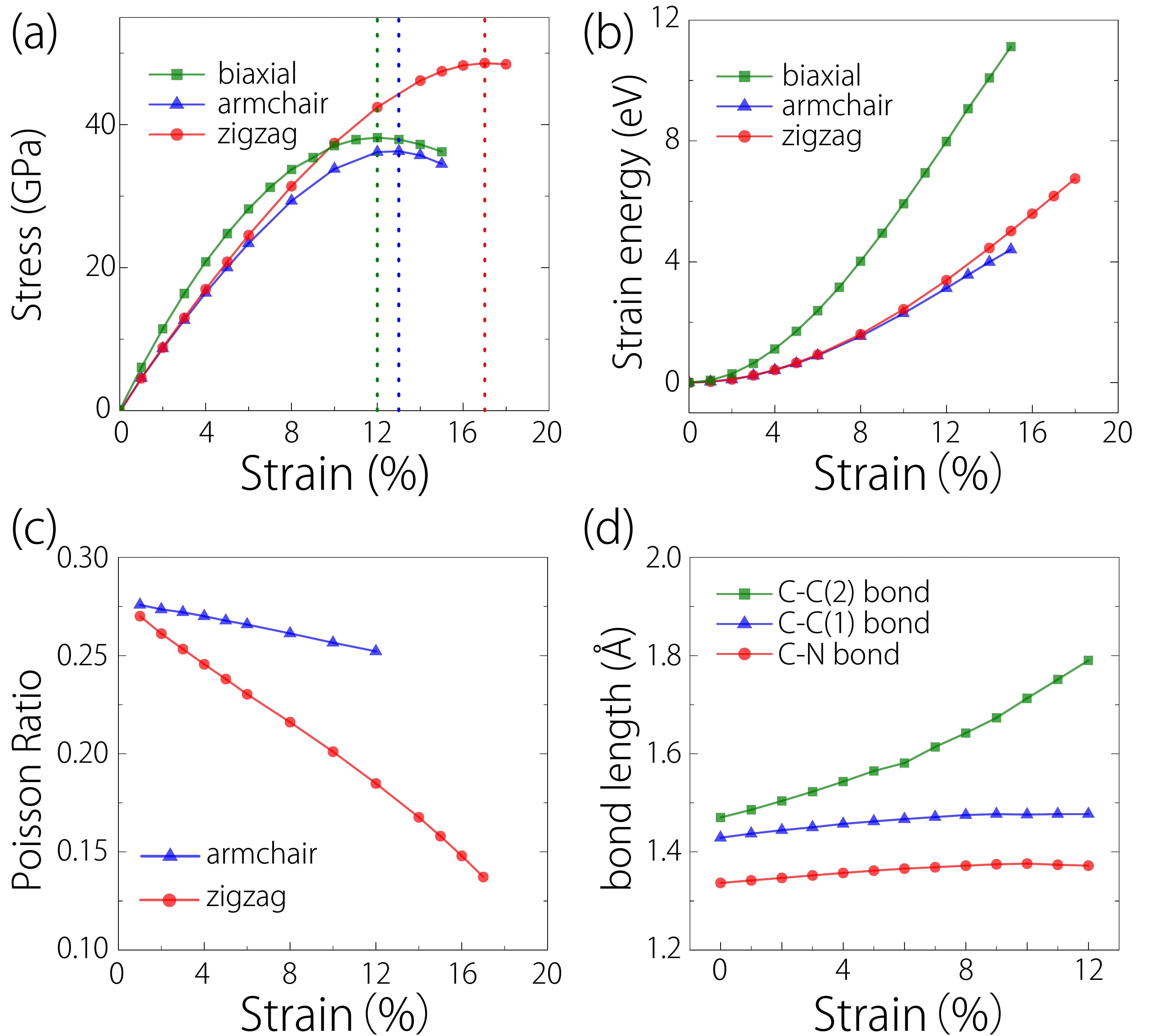}
\caption{(a) Strain-stress relations for monolayer C$_2$N-$h$2D with different types of strain. The vertical dashed lines indicate the critical tensile strains (up to 12\%, 13\%, and 17\% for biaxial strain and uniaxial strain in the armchair and zigzag
directions, respectively). (b) Strain energy $E_S$ as a function of strain (for the three types of strain). (c) Poisson ratio $\nu$ as a function of uniaxial strain in the zigzag and armchair directions, respectively. (d)  The variation of the three bond-lengths versus applied biaxial strain.}
\label{fig2}
\end{figure}

First, to estimate the elastic limit of C$_2$N-$h$2D, we calculate the strain-stress relations.\cite{roun2001,luo2002} The results are presented in Fig.~\ref{fig2}(a). One observes that C$_2$N-$h$2D exhibits linear elastic response until a strain of about 8\%. While the material can sustain a biaxial strain up to 12\%, the uniaxial tensile strain limits are 13\% and 17\% along the armchair and zigzag directions, respectively. These values are smaller than those for graphene and hexagonal boron nitride due to the presence of holes and relatively weaker C-N bonds,\cite{kim2009,lee2008,wu2013} nevertheless, they are still quite large compared with conventional 3D materials. One also notices that strain is more easily applied along the armchair direction than the zigzag direction. For small deformations, the elastic properties of 2D materials can be characterized by the in-plane stiffness constant, defined as $C= (1/S_0)(\partial^2 E_S/\partial \varepsilon^2)$,\cite{wang2014} where $S_0$ is the equilibrium area and the strain energy $E_S$ is given by the energy difference between the strained and unstrained systems. We plot the strain energy as a function of strain strength in Fig.~\ref{fig2}(b), from which the typical quadratic dependence can be observed at small deformations. The calculated stiffness constants along armchair and zigzag directions have similar values about $71$ N/m. Such values are significantly smaller than that of graphene ($\sim 340\pm 40$ N/m),\cite{lee2008} MoS$_2$ (140 N/m),\cite{Peng2013} and BN (267 N/m),\cite{topsakal2010} showing that C$_2$N-$h$2D is much softer, which would facilitate the strain engineering of its properties.

For uniaxial strains, a tensile strain applied in an axial direction generally results in a compression in the transverse direction. This behavior is reflected by the Poisson ratio, here defined as $\nu=-\varepsilon_\text{transverse}/\varepsilon_\text{axial}$. Fig.~\ref{fig2}(c) shows the results of Poisson ratios for the two types of uniaxial strain. One finds that in the small strain limit ($\varepsilon_\text{axial}\rightarrow 0$), the Poisson ratios for the two directions converge to the same value $\approx0.28$, showing the response is isotropic at small strains. However, for larger strains, their values decrease monotonically at different speeds with the applied strain, manifesting the anisotropy in the crystal structure. Compared with graphene ($\nu\sim0.16$ under small strains\cite{gui2008,cheng2011}), the Poisson ratios of C$_2$N-$h$2D are much larger, which can be expected due to its holey structure. In addition, we note that at large strains $\nu_{a}>\nu_{z}$ (here and hereafter, subscripts $a$ and $z$ stand for armchair and zigzag directions, respectively) for monolayer C$_2$N-$h$2D.

\begin{figure}
\includegraphics[width=9cm]{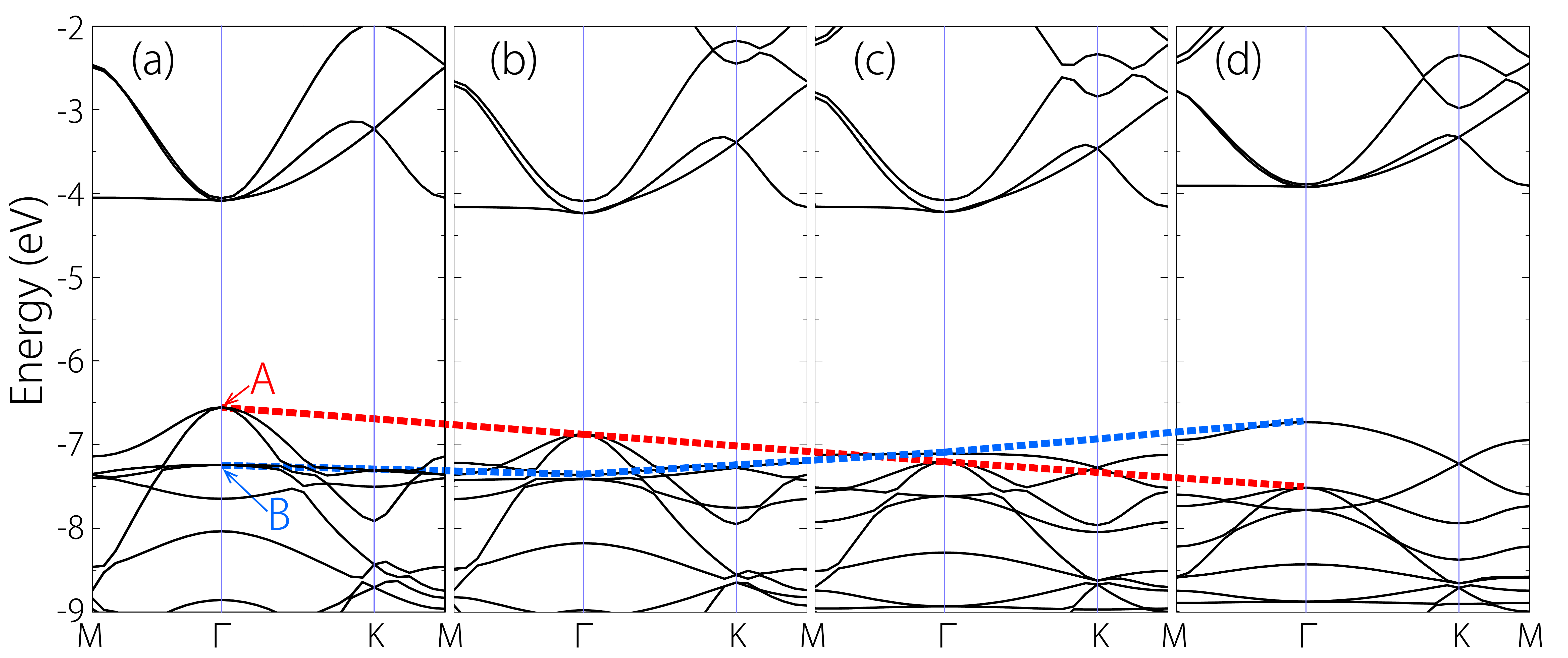}
\caption{Band structures of C$_2$N-$h$2D with biaxial strains (a) $\varepsilon=0\%$, (b) $\varepsilon=4\%$, (c) $\varepsilon=8\%$, and (d) $\varepsilon=12\%$. Besides the change in bandgap size, the band ordering around VBM is switched for $\varepsilon\lesssim 8\%$. The dashed lines are guide for eye for
the energy shifts of states A and B. The energy is referenced to the vacuum level.}
\label{fig3}
\end{figure}

Then we analyze the strain effects on the electronic band structures C$_2$N-$h$2D. The calculated band structure in the absence of strain is shown in Fig.~\ref{fig3}(a). It shows a semiconductor with a direct bandgap at $\Gamma$ point about 2.46 eV, which compares well with the reported optical bandgap and agrees with other theoretical studies.\cite{mahm2015,zhang2015b} Around the conduction band minimum (CBM), there are three bands almost degenerate at $\Gamma$ point. The lowest conduction band is quite flat along $\Gamma$-$M$. Around VBM, there are two bands with different dispersions (heavy-hole and light-hole) but degenerate at $\Gamma$ point. 

\begin{figure}
\includegraphics[width=8.8cm]{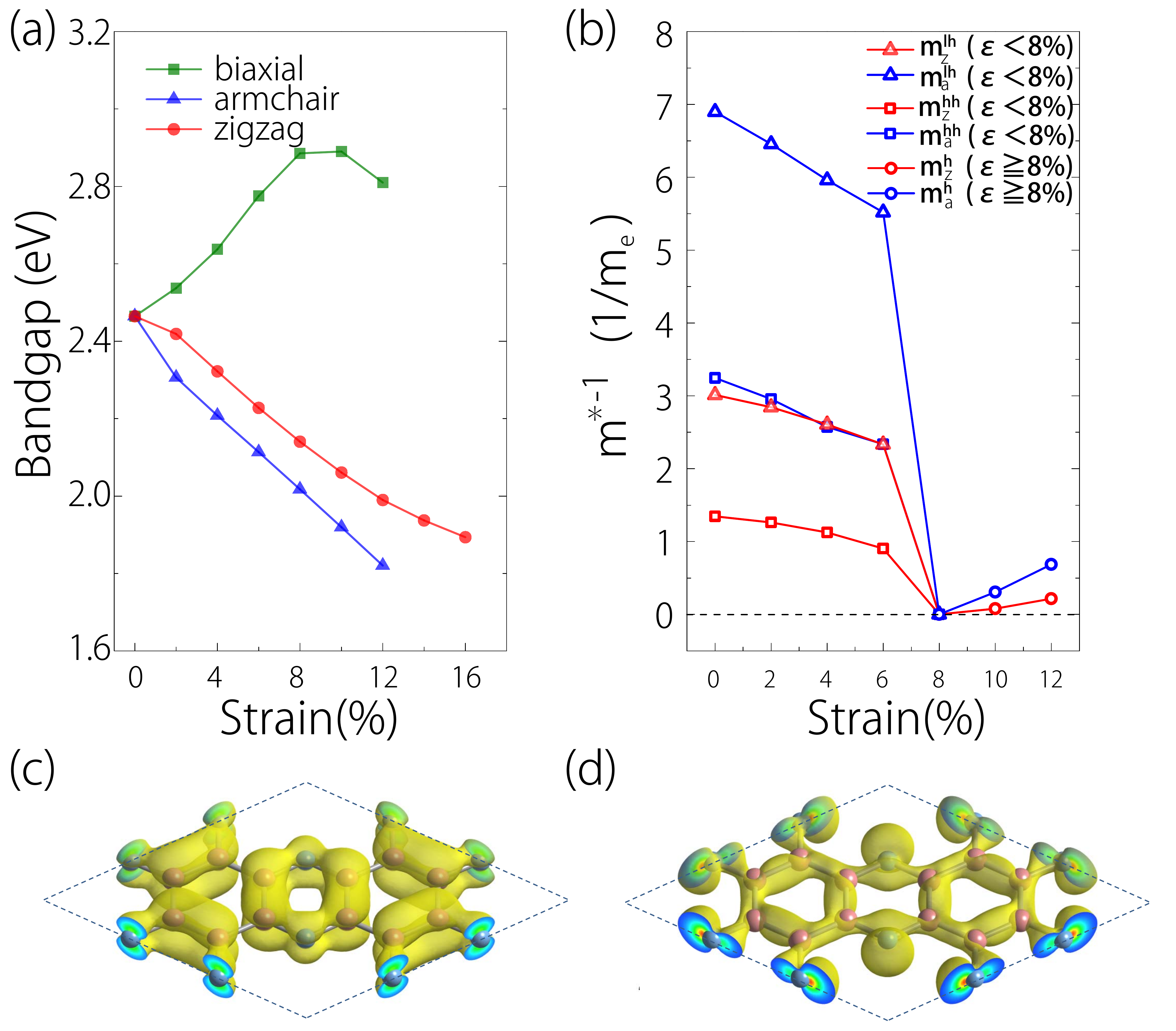}
\caption{(a) Bandgap of monolayer C$_2$N-$h$2D as a function of the three types of strain. (b) Effective mass (plotted with $1/m^*$) of hole carriers versus  biaxial strain. For $\varepsilon< 8\%$, there are two types of holes: heavy hole and light hole. Drastic change occurs around $8\%$ strain where band crossing occurs. And for $\varepsilon\geq 8\%$, there is only one type of hole with a large effective mass. The colors red and blue correspond to zigzag and armchair directions, respectively. (c, d) are the charge density contour plots for (c) A and (d) B states at $\Gamma$ point, as indicated in Fig.~\ref{fig3}.}
\label{fig4}
\end{figure}

The band structures of C$_2$N-$h$2D under biaxial strains $\varepsilon=4\%$, $8\%$, and $12\%$ are shown in Fig.~\ref{fig3}(b)-(d). One observes that while the system remains a direct-gap semiconductor in the considered range, the bandgap size is modified by strain. As shown in Fig.~\ref{fig4}(a), the bandgap increases from 2.46 eV to about 2.89 eV around a strain of 8\%.  Interestingly, above 8\% strain, the gap size starts to decrease (with a value $\sim2.81$ eV at $\varepsilon=12\%$).
This non-monotonic dependence can be understood by a closer look at Fig.~\ref{fig3}. The CBM energy has little change with strain. In contrast, the bands around VBM are strongly affected by strain: the heavy hole and light hole bands at the original VBM (marked by point A) are shifted down in energy leading to an increase of bandgap; meanwhile, another valence band initially below the VBM by about 1 eV (marked by point B) is moved up with increasing strain. Close to $\varepsilon=8\%$, A and B cross each other and the band ordering is reversed. After this transition, B becomes the new VBM and keeps moving up in energy, thereby turning the trend of bandgap change from increasing to decreasing.

To better understand this interesting strain-induced band crossing, we analyze the character of the relevant electronic states. In Fig.~\ref{fig4}(c)-(d), we plot the charge density distributions corresponding to A and B, respectively. One can observe that the states at A mainly consist of $p_z$-orbitals of C and N atoms, while B is mainly composed of $p_x$ and $p_y$-orbitals. More importantly, A shows anti-bonding character along the C-C(2) bonds, over which B instead shows the bonding character. In Fig.~\ref{fig2}(d), we plot the variation of the three bond-lengths versus the biaxial strain. It clearly shows that the C-C(2) bonds are most sensitive to strain. The lattice expansion can be mostly attributed to the stretching of C-C(2) bonds. Therefore, due to the different bonding characters along C-C(2), A and B respond differently to strain: A is pulled down while B is pushed up.

\begin{figure}
\includegraphics[width=9cm]{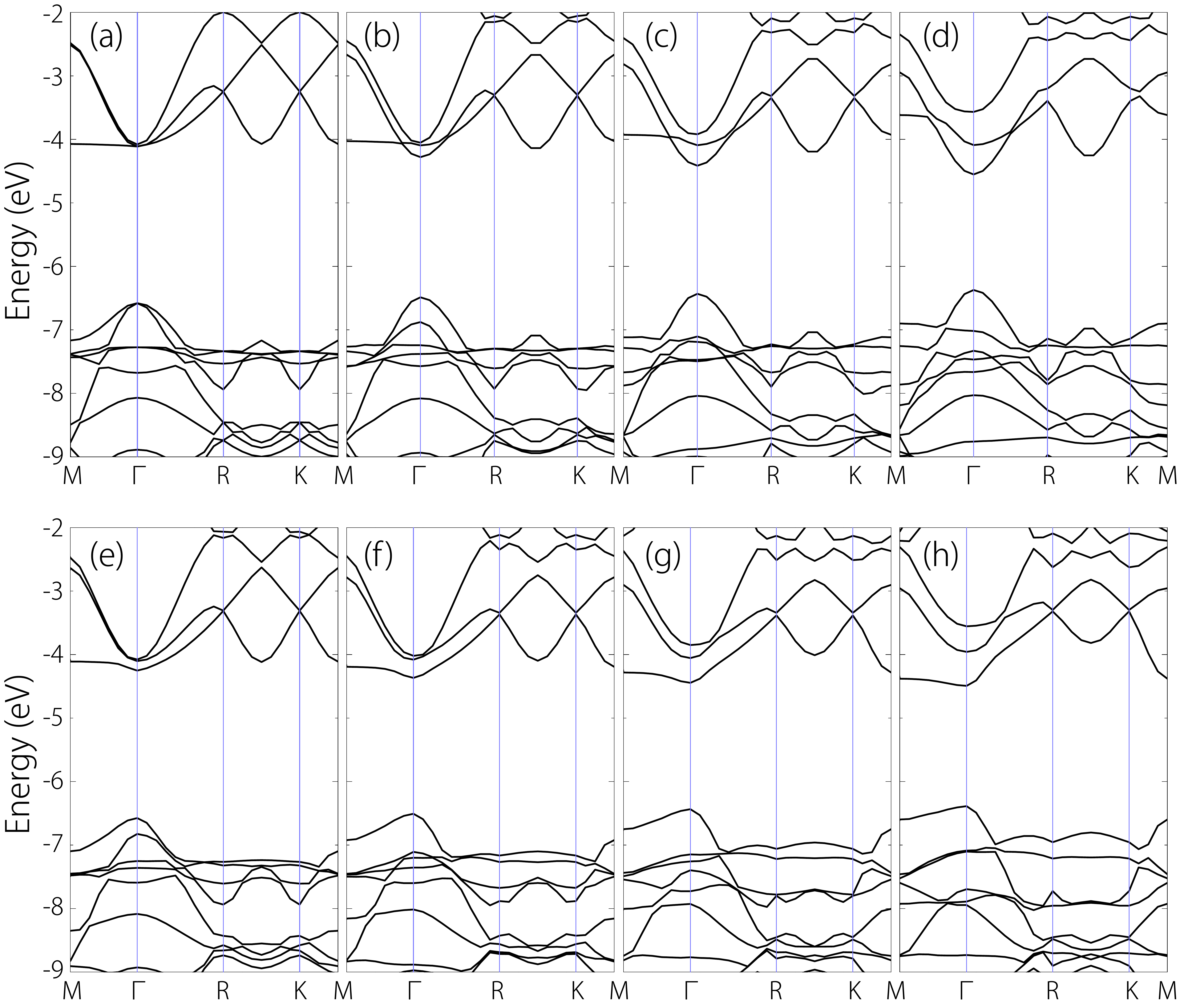}
\caption{ Band structures of C$_2$N-$h$2D under uniaxial strians: (a) no strain, (b) $\varepsilon_a=4\%$, (c) $\varepsilon_a=8\%$,  (d) $\varepsilon_a=12\%$, (e) $\varepsilon_z=4\%$, (f) $\varepsilon_z=8\%$, (g) $\varepsilon_z=12\%$, and (h) $\varepsilon_z=16\%$. Here $a$ and $z$ label the armchair and zigzag directions, respectively. }
\label{fig5}
\end{figure}

The switch of band ordering changes the character of VBM, which is expected to have strong effects on the system properties. In particular, we note that the band with B is quite flat. Hence upon the band crossing, there will be a dramatic increase in the hole effective mass. In Fig.~\ref{fig4}(b), we plot the inverse of hole effective mass $1/m^*_i=(1/\hbar^2)|\partial^2 E/\partial k^2_i|$ ($i=a,z$ labels the armchair and zigzag directions) as a function of strain. Before band crossing ($\varepsilon<8\%$), there are two types of hole carriers, heavy hole (hh) and light hole (lh), with different effective masses $m^{hh}_i$ and $m^{lh}_i$. One observes that for both hole bands, $m_a$ is about half of $m_z$, and they increase slightly with strain. Indeed, close to 8\% strain, there is a dramatic increase in the effective mass due to the band crossing. The inverse of effective masse is close to zero for $\varepsilon\approx8\%$ along both directions. This jump in effective mass could potentially quench the hole transport. It hence suggests the possibility to control the hole transport using strain.

\begin{figure}
\includegraphics[width=8.6cm]{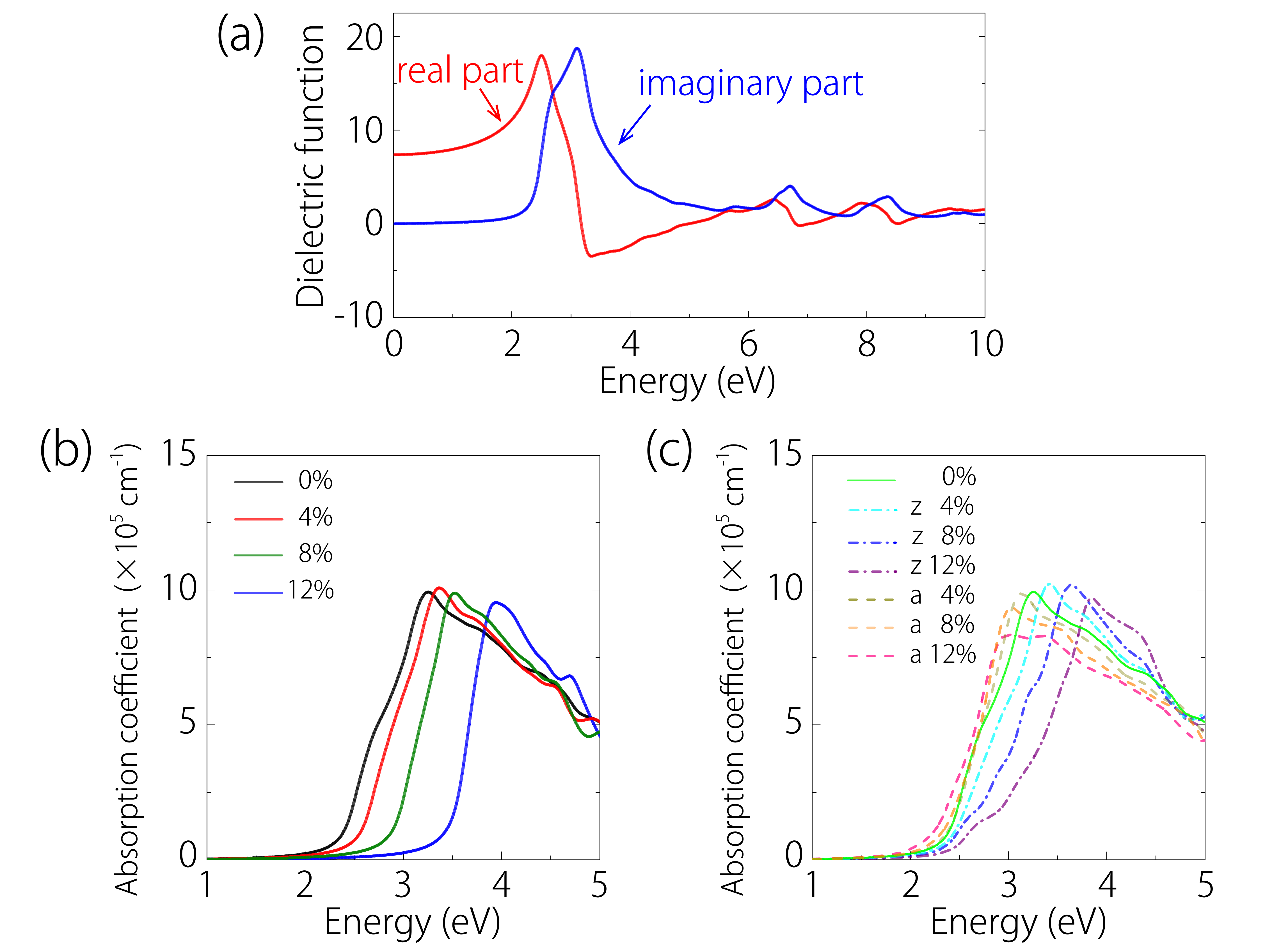}
\caption{(a) The real and imaginary parts of the dielectric function for unstrained monolayer C$_2$N-$h$2D. (b) and (c) show the strain effects on the optical absorption coefficient for (b) biaxial strain and (c) uniaxial strain.}
\label{fig6}
\end{figure}

Next, we consider the effects of uniaxial strain. The resulting band structures for a few representative strains are shown in Fig.~\ref{fig5}. The system remains a semiconductor with direct bandgap at $\Gamma$ point. However, different from the case of biaxial strain, the gap monotonically decreases for both $\varepsilon_a$ and $\varepsilon_z$ . This bandgap variation is illustrated in Fig.~\ref{fig4}(a). The decrease of gap size can be up to 0.6 eV at $\varepsilon_a=12\%$. The average effective masses for both electrons and holes are decreased by strain. In addition, from the band structures in Fig.~\ref{fig5}, we notice that due to the breaking of $C_3$ symmetry by strain, the originally degenerated states at both CBM and VBM are split up in energy.

The discussions above demonstrate that strain is an effective approach for tuning the band structures of C$_2$N-$h$2D monolayers. With different strains, the gap can be tuned in a wide range from about 1.8 eV to 2.9 eV. Since the material maintains a direct bandgap in the visible light spectrum under stain, it is naturally expected to have promising applications for optoelectronic devices. In Fig.~\ref{fig6}(a), we plot the real ($\epsilon_1$) and imaginary ($\epsilon_2$) parts of the dielectric function $\epsilon(\omega)$ for the unstrained system. The curves show typical features of dielectric functions for insulators. The first peak of $\epsilon_2(\omega)$ is around 3 eV, where the real part changes sign. From the dielectric function, we can obtain the optical absorption coefficient $\alpha(\omega)=\omega \epsilon_2/(cn)$, where $c$ is the speed of light, and $n=(\sqrt{\epsilon_1^2+\epsilon_2^2}+\epsilon_1)^{1/2}/\sqrt{2}$ is the index of refraction. In Fig.~\ref{fig6}(b) and \ref{fig6}(c), we plot the absorption coefficient for biaxial and uniaxial strains, respectively. One observes that the absorption for unstrained case has a broad peak around 3 eV with a maximum value as large as $10^6$ cm$^{-1}$. The absorption profile including the peak position can be effectively tuned by strain, resulting in a large absorption in a wide range from 2.5 eV to 5 eV. The value of $\alpha$ is comparable to that of graphene, MoS$_2$ or phosphorene ($\lesssim 10^6$ cm$^{-1}$ in the corresponding range).\cite{nair2008,mak2010,wang2012,qiao2014} Since C$_2$N-$h$2D is of only one atom thickness, thinner than MoS$_2$ and phosphorene, it will have great potential for compact and flexible optoelectronic devices.

In summary, based on first-principles calculations, we demonstrate that strain can effectively tune the electronic and optic properties of the newly discovered C$_2$N-$h$2D material. The material can sustain a large strain $\geq 12\%$ and has a small stiffness constant. The bandgap remains direct under strain and can be tuned in a wide range from 1.8 eV to 2.9 eV. For biaxial strain, an interesting band crossing effect occurs for VBM at $\varepsilon\lesssim 8\%$. This changes the character of hole carriers and especially leads to a dramatic increase of the hole effective mass. With strain, large optical absorption can be achieved in a wide range from 2.5 eV to 5 eV. Our results suggest effective methods to engineer the C$_2$N-$h$2D material for future electronic and optoelectronic device applications.

 The authors thank Jin Li and D. L. Deng for helpful discussions. This work was supported by the MOST Project of China (Grants No.2014CB920903), SUTD-SRG-EPD2013062, and the NSFC (Grants No.11225418 and No.11574029). We acknowledge computational support from the National Supercomputer Center in Tianjin and Texas Advanced Computing Center.

\bibliographystyle{apsrev4-1}

\begin{thebibliography}{100}

\bibitem{novo2004} K. S. Novoselov, A. K. Geim, S. V. Morozov, D. Jiang, Y. Zhang, S. V. Dubonos, I. V. Grigorieva, A. A. Firsov, Science {\bf 306}, 666 (2004).
\bibitem{geim2007} A. K. Geim and K. S. Novoselov, Nat. Mater. {\bf 6}, 183 (2007).
\bibitem{xu2013} M. Xu, T. Liang, M. Shi and H. Chen, Chem. Rev. {\bf 113}, 3766 (2013).
\bibitem{butl2013} S. Z. Butler, S. M. Hollen, L. Cao, Y. Cui, J. A. Gupta, H. R. Guti\'errez, T. F. Heinz, S. S. Hong, J. Huang, A. F. Ismach, E. Johnston-Halperin, M. Kuno, V. V. Plashnitsa , R. D. Robinson, R. S. Ruoff, S. Salahuddin, J. Shan, L. Shi, M. G. Spencer, M. Terrones, W. Windl and J. E. Goldberger, ACS Nano {\bf 7}, 2898 (2013).

\bibitem{mahm2015} J. Mahmood,	E. K. Lee,	M. Jung,	D. Shin,	I.-Y. Jeon,	S.-M. Jung,	 H.-J. Choi,	J.-M. Seo,	 S.-Y. Bae,	S.-D. Sohn,	N. Park,	J. H. Oh,	H.-J. Shin	and J.-B. Baek, Nat. Commun. {\bf 6}, 6486 (2015).
\bibitem{zhang2015} R. Zhang, B. Li, and J. Yang, Nanoscale {\bf 7}, 14062 (2015).
\bibitem{zhang2015b} R. Zhang and J. Yang, arXiv:1505.02768.
\bibitem{xu2015} B. Xu, H. Xiang, Q. Wei, J. Q. Liu, Y. D. Xia, J. Yin and Z. G. Liu, Phys. Chem. Chem. Phys. {\bf 17}, 15115 (2015).


\bibitem{kim2009} K. S. Kim, Y. Zhao, H. Jang, S. Y. Lee, J. M. Kim, K. S. Kim, J.-H. Ahn, P. Kim, J.-Y. Choi and B. H. Hong, Nature {\bf 457}, 706 (2009).
\bibitem{lee2008} C. Lee, X. Wei, J. W. Kysar, J. Hone, Science {\bf 321}, 385 (2008).


\bibitem{cast2012} A. Castellanos-Gomez, M. Poot, G. A. Steele, H. S. van der Zant, N. Agrait, and G. Rubio-Bollinger, Nano. Res. Lett. {\bf 7}, 233 (2012).
\bibitem{bert2011} S. Bertolazzi, J. Brivio, and A. Kis, ACS Nano {\bf 5}, 9703 (2011).
\bibitem{peng2014} X. Peng, Q. Wei, and A. Copple, Phys. Rev. B {\bf 90}, 085402 (2014).


\bibitem{levy2010} N. Levy, S. A. Burke, K. L. Meaker, M. Panlasigui, A. Zettl, F. Guinea, A. H. Castro Neto, and M. F. Crommie, Science {\bf 329}, 5991 (2010).
\bibitem{guin2010} F. Guinea, M. I. Katsnelson, and A. K. Geim, Nat. Phys. {\bf 6}, 30 (2010).
\bibitem{feng2012} J. Feng, X. Qian, C.-W. Huang, and J. Li, Nat. Photon. {\bf 6}, 866 (2012).
\bibitem{zhang2013a} Q. Y. Zhang, Y. C. Cheng, L. Y. Gan, and U. Schwingenschl\"ogl, Phys. Rev. B {\bf 88}, 245447  (2013).
\bibitem{rodi2014} A. S. Rodin, A. Carvalho, and A. H. Castro Neto, Phys. Rev. Lett. {\bf 112}, 176801 (2014).
\bibitem{fei2014} R. Fei and L. Yang, Nano Lett. {\bf 14}, 2884 (2014).

\bibitem{vasp1} G. Kresse and J. Furthmuller, Phys. Rev. B {\bf 54}, 11169 (1996).
\bibitem{vasp2} G. Kresse and J. Furthmuller, Comput. Mater. Sci. {\bf 6}, 15 (1996).
\bibitem{paw1}  P. E. Blochl, Phys. Rev. B {\bf 50}, 17953 (1994).
\bibitem{paw2}  G. Kresse and D. Joubert, Phys. Rev. B {\bf 59}, 1758 (1999).
\bibitem{pbe}   J. P. Perdew, K. Burke, and M. Ernzerhof, Phys. Rev. Lett. {\bf 77}, 3865 (1996).
\bibitem{hse1}   J. Heyd, G. E. Scuseria, and M. Ernzerhof, J. Chem. Phys. {\bf 118}, 8207 (2003).
\bibitem{hse2}   J. Heyd, G. E. Scuseria, and M. Ernzerhof, J. Chem. Phys. {\bf 124}, 219906 (2006).


\bibitem{roun2001} D. Roundy and M. L. Cohen, Phys. Rev. B {\bf 64}, 212103 (2001).
\bibitem{luo2002} W. Luo, D. Roundy, M. L. Cohen, and J. W. Morris Jr., Phys. Rev. B {\bf 66}, 094110 (2002).
\bibitem{wu2013} J. Wu, B. Wang, Y. Wei, R. Yang and M. Dresselhaus, Mater. Res. Lett. {\bf 1}, 200 (2013).

\bibitem{wang2014} G. Wang, M. Si, A. Kumar and R. Pandey, Appl. Phys. Lett. {\bf 104}, 213107 (2014).
\bibitem{Peng2013} Q. Peng and  S. De, Phys. Chem. Chem. Phys. {\bf 15}, 19427 (2013).
\bibitem{topsakal2010} M. Topsakal, S. Cahangirov, and S. Ciraci,  Appl. Phys. Lett. {\bf 96}, 091912 (2010).

\bibitem{gui2008} G. Gui, J. Li, and J. Zhong, Phys. Rev. B {\bf 78}, 075435 (2008).
\bibitem{cheng2011} Y. C. Cheng, Z. Y. Zhu, G. S. Huang, and U. Schwingenschl\"{o}gl, Phys. Rev. B {\bf 83}, 115449 (2011).

\bibitem{nair2008}    R. R. Nair, P. Blake, A. N. Grigorenko, K. S. Novoselov, T. J. Booth, T. Stauber, N. M. R. Peres, and A. K. Geim, Science {\bf 320}, 1308 (2008).
\bibitem{mak2010} K. F. Mak, C. Lee, J. Hone, J. Shan, and T. F. Heinz, Phys. Rev. Lett. {\bf 105}, 136805 (2010).
\bibitem{wang2012} Q. H. Wang,	K. Kalantar-Zadeh,	A. Kis,	J. N. Coleman	and M. S. Strano, Nat. Nanotech. {\bf 7}, 699 (2012).
\bibitem{qiao2014} J. Qiao, X. Kong, Z.-X. Hu, F. Yang and W. Ji, Nat. Commun. {\bf 5}, 4475 (2014).

\end{thebibliography}

\end{document}